# SPIRAP – Wireless Uplink Random Access Protocol Using Spinal Code

Dan Raphaeli and Snir Nisim Tel Aviv University, Israel

*Abstract—* **In this paper we present SPIRAP - SPinal Random-Access Protocol, a new method for multiuser detection over wireless fading channel. SPIRAP combines sequential decoding with rateless Spinal code. SPIRAP appears to be an efficient protocol for transmitting small packets in a minimally controlled network and can be attractive for the Internet of Things (IOT) applications. The algorithm applies equivalently to any other rateless code, we chose Spinal for its good performance. We show that SPIRAP may achieve higher rate than TDMA and ALOHA in some cases and without the need for user's synchronization.**

## I. INTRODUCTION

Multiple access protocols over random wireless channels have a great importance in many communications applications, some of them extremely widespread like Wi-Fi, Bluetooth and Zigbee. New emerging set of Internet of Things (IOT) applications require support for a multitude of nodes in random varying fading channels, each transmitting a very short packet at random times without any synchronization.

In general, assume N mobile stations (MS) communicating with a Base Station (BS) simultaneously and asynchronously over Additive White Gaussian Noise (AWGN) and fast Rayliegh fading channel, with a very small amount of data for each packet. In such systems, the Signal to Noise (SNR) for each user is unknown and time varying. The transmitter hence needs to know the channel conditions and accordingly choose the best code rate and modulation in every transmission for each user. An additional large issue is multiple access. How to let the users concurrently use the common RF channel in an efficient way.

For the first problem, adaptive rate protocols are commonly used. These protocols suffer from difficulty to distinguish between channel conditions and collisions and require that the channel conditions will change slow enough and sufficient traffic to estimate the channel conditions.

Multiple access protocols may be divided into three main families as described in Figure 1. The first family comprises the random-access protocols, with minimum central control. These protocols are designed to avoid or resolve collisions when two or more users wish to transmit at the same time. The second family includes controlled access protocols like NORA (Non-Orthogonal Random Access), which exercise and handshakes, in which the base station identifies each user and commands it's transmission power in accordance with the specific channel conditions, as described hereinafter. The third family of protocols relies on channelization protocols as TDMA, FDMA, or CDMA.

One of the most used protocol is the IEEE802.11 (Wi-Fi), a member of the random-access protocol's family. It applies Carrier Sense Multiple Access (CSMA), which means listening and waiting for a clear channel before transmitting. Another well-known random-access protocol is ALOHA, in which collisions are not avoided but resolved by retransmissions after random back-off times.

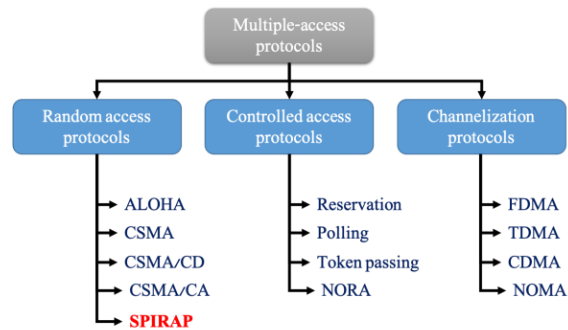

Figure 1: multiple access protocols

As mentioned, wireless protocols have a lot of overhead. In a controlled network most of the overhead occur when establishing the connection, but once a timeslot is assigned, there is no neeFd to pass information in each packet, except signals that are used for channel estimation and synchronization. In an uncontrolled network as a CSMA based, an additional overhead must be calculated for the operation of the protocol. Thus, each packet includes PHY header and MAC header. The PHY header is used for synchronization and parameter estimation, and the MAC header for handling information about the source address, the coding and modulation rate in the data part and another field is required for the operation of the collision avoidance protocol. Since the MAC header must be parsed in any SNR condition, it should be transmitted at the lowest rate supported by the protocol and protected by a large CRC. When the number of data bits in each packet is low, and the SNR condition is good, allowing large rate, this overhead can be very significant [8], and may climb up to 50% of the raw data rate.

### A. Channel Coding

Two of the highest performance error correction codes are the Low Density Parity Check (LDPC) code of Gallager [1] and the Turbo codes [2]. These two channel codes are shown to operate at rates very close to the channel capacity for AWGN channels. They both have efficient decoding algorithms. hence, they are efficiently implemented in a variety of communication standards. The codes mentioned above are fixed-rate codes, that is, they take k input symbols and generate n coded symbols with a fixed-rate of n/k. The rate of the fixed-rate codes is determined according to the communication channel's conditions either at the design or at the instant of communication. The communication protocols using these codes must have information about the channel conditions continuously at the transmitter side, so that the transmitter can adapt its coding rate to the environment conditions. This disadvantage is solved by a relatively new type of codes - rateless codes.

A rateless code is defined as a channel code in which the higher rate codes are prefixes of the lower rate codes. The transmitter using a rateless encoder generates encoded symbols until the receivers acknowledge the transmission is successful or a timeout for transmission time is reached. This makes the rateless transmission's coding rate subject to change at each transmission. In other words, a higher code rate is possible when the channel conditions are good and a lower rate when the channel conditions are bad. Since the coding rate is not fixed ahead of transmission, the coding process adapts itself almost perfectly to the changing channel conditions. Rateless codes offer an alternative to rate adaptation. Rate adaptation can work only in slow channel variations and it is difficult to distinguish between thermal noise related packet loss and collisions.

There are few problems using reateless codes within exsiting protocols. ALOHA needs a mechanism of detecting collisions at the reciever, this may be a difficult tast when using rateless codes since the reciever has to distinguish a collision from a simple noisy packet that cannot yet be decoded. In CSMA, we have the same problem as in ALOHA, when there is a need to resolve collisions using ACK since there is no ACK until the final pass is transmitted. Furthermore, CSMA requires the MAC header to be transmitted in a low rate, limitting the range of rates that can be supported, and cause large overhead .

In this paper, we propose a multiple access protocol integrated with a channel code, that is useful for arbitrary network with multiple users, unknown and fast fading channel, with short packets. We called it SPinal Random Access Protocol (SPIRAP). This protocol is based on using a rateless code to decode multi-user in a multiple access scenarios. Specificly we propose the use of Spinal code [3]. In many conditions we show higher total rate than TDMA, CSMA or ALOHA, when all use the same rateless code.

SPIRAP includes a sequential decoding process. In case of a collision, first the strongest user is decoded while treating all the rest of the users as noise. After a successful decoding of a user, the user is subtracted and the next strongest user is decoded and so on. There is some similiarity between SPIRAP and the recently introduced NORA[7]. But there are profound differences between the two as further discussed hereinafter.

Since there is no knowledge about how many users are transmitting in the current timeslot, an input power threshold at the decoder is being used. While the threshold is crossed the decoder keeps trying to decode more and more passes and sequentialy substract the decoded ones. This way it is redundent for each user to send his MAC address and as can be seen later, we evaluate gain and phase estimators through the decoding process (assuming flat channel), so that there is no need for pilot tones.

Using this method the MAC header is dismissed and the preamble overhead is reduced resulting a more efficient link for small messages. in addition to the reduction of the overhead, rulingout the MAC header helps solving a major problem when rateless code is used as only at the end of the session the data bits can be resolved. This timing is too late for maintaining a MAC protocol like 802.11 in which all users must decode a MAC header in each packet.

Moreover, in SPIRAP there is no need for the users to know or sync with each other or even "hear" each other (that case is not available in common CSMA approaches). This capability solves one of the problems found in CSMA – the hidden node problem. Solving the hidden node problem requires some overhead of the RTS-CTS mechanism.

SPIRAP has two main disadvantages; Since there is no header, we need to estimate the channeland when a user started transmitting.

## II. SPINAL CODE

A recently proposed rateless code is the rateless spinal code [3]. The spinal codes generate encoding symbols by sequentially applying a hash function to smaller portions of the message to be transmitted. Applying the hash function creates encoded symbols very different from each other even when the two messages differ only in a single bit. This in turn makes spinal codes robust to noise and interference effects. In [3], it is shown through simulations that the spinal decoder results better performance than Raptor, LDPC and Strider, former rateless codes, at a large range of SNR values and in fading channels [4].

Spinal codes achieve close to capacity rate values for a large range of SNR values as can be seen in Figure 1 from [3].

Using a hash function to generate encoded symbols makes the decoding of spinal codes harder by inverting the encoder structure. As described in [3], an approximate maximum-likelihood decoder, called Bubble-decoder, is used to efficiently decode spinal codes. The Bubble decoder traverses the tree out of possible transmitted symbols.

While traversing the tree, the decoder prunes the least likely paths so that the decoding operation can be completed with reasonable complexity instead of the exponential complexity of the maximum-likelihood decoder.

## III. SYSTEM MODEL

A. General

Assume N Mobile stations (MS) communicating with a Base station (BS). A feedback channel from BS to all MS containing an acknowledge packet (ACK) which indicates to any specific MS if the reception of its last message was successful or not.

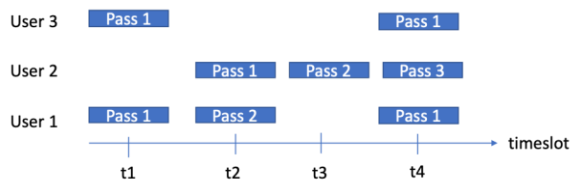

Figure 2 - SPIRAP timeslots illustarion

All MS are synchronized by the BS and the data from MS to BS is organized in constant timeslots. In each timeslot each MS may transmit a packet according to a Poisson distribution with parameter $\gamma$. Each packet that is transmitted in a timeslot $i$, from MS to BS contains $m$ QAM symbols $x_{ij}$. The symbols are transmitted over arbitrary flat fading channel

$$r_{ij} = \sum_{n=1}^{N} \alpha_i^n x_{ij}^n + w_{ij} \quad (1)$$

where $\alpha_i^n$ is some fading process. We assume that the fading is constant over one slot. The model used in this paper is Rayleigh fading + lognormal process. The processes are generated in simulation by filtering independent fading process, as can be seen in Figure 3, $\alpha_i^n$, is independent from MS to MS, and $w_{ij} \sim CN(0, N_0)$ is additive Gaussian noise. Generally using SPIRAP, any Mobile Station (MS) can send data any time, no fixed time to transmit, no fixed sequence of stations. Each MS can start the transmission at any slot (without carrier sense), it keeps re-transmitting more passes from the same packet until the BS signals it that it is decoded successfully or it gives up (timeout) and the packet is discarded (or retransmitted). Unless stated otherwise, new packets are sent, after ACK or timeout, using Poisson distribution with parameter $\gamma$.

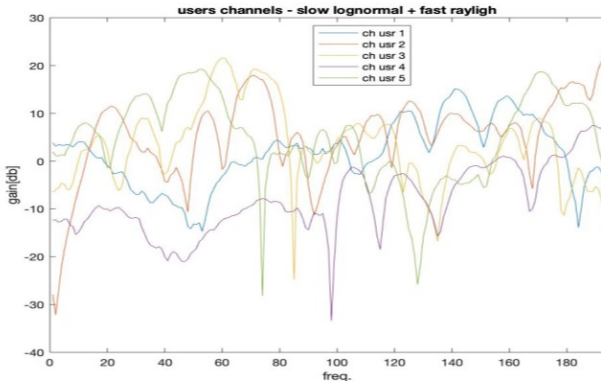

Figure 3: 5 users channel illustration fast Rayligh + slow lognormal fadding

### B. Concept Essentials

Since the SNR in a single pass is usually insufficient for decoding, the MAC header cannot be decoded, thus we need to avoid pilots coliding users. SPIRAP protocol is based on the following elements:
- No carrier sense, no tx control, transmit at will
- collisions are allowed and resolved.
- Decode the strongest user, substract it on success and retry the rest of the users
- There is no MAC header, BS tries to decode each user in each pass.
- When a user is being decoded, other currently transmitting users are treated as noise relative to that user.
- For simplicity of the simulation, users are slot synchronized, and transmits the same sized packets.
- Use of spinal rateless code (or any other rateless code)
- Resolve collisions by rateless code re-transmissions where interfering users are treated as noise.
- No pilots are used, Gain and phase estimators are based on data only.
- Gain and phase are independent between passes
- Packet header is not transmitted, existence of any transmission is detected by a power threshold.
- False detection of power threshold is not detrimental. It just might add false pass to active user passes and cause unnecessary computations.
- Miss detection of power threshold is not detrimental. it just will reduce the rate due to additional passes required for recovery.

### IV. SPIRAP ALGORITHM DESCRIPTION

The SPIRAP algorithm is shown in Algorithm 1.

*Algorithm 1: The SPIRAP Algorithm*

```
for Each timeslot i
  if (var(r_ij) > N0+Start_Th)
    if ~startFlag    % so this is the first time we got start
      StartTimeSlot=i;
      startFlag=true;
    end
    No_signal_cnt=0
  else
    No_signal_cnt++
    if No_signal_cnt>No_signal_th
      startFlag=false;
    end
  end
  if startFlag
    while(1)
      if Decoded(n)
        StartTimeSlot(n) = i+1
        Continue;
      end
      Decoded(n) = SPIRAPDecoder(r, (StartTimeSlot(n)));
      if Decoded(n)
        r(StartTimeSlot(n):i) = r(StartTimeSlot(n):i)-
          g*x_i(StartTimeSlot(n):i)
        for i= StartTimeSlot(n):i
          if (var(r(i)) > N0+Start_Th)
            break;
          else
            n++;
          end
        end
        for m=1:n
          if StartTimeSlot(m)==StartTimeSlot(n)
            StartTimeSlot(m)=i;
          end
        End
        n=1;
      end
      if max decoding attempts reached
        break;
      end
    end
    Ask for new pass for each user with Decoded (n)==0
  end
```

## A. Gain Estimation

We have added a gain estimator since each user in each slot has a random gain. Assuming only one user is transmitting is a good approximation to the case of one strong user. Obviously if the user powers are even, the estimator will be inaccurate. This gain estimation error will cause some degradation in the final algorithm performance. The estimator formula is: We have added a gain estimator since each user in each slot has a random gain. Assuming only one user is transmitting is a good approximation to the case of one strong user. Obviously if the user powers are even, the estimator will be inaccurate. This gain estimation error will cause some degradation in the final algorithm performance. The estimator formula is:

$$|\hat{\alpha}_i| = \sqrt{\frac{VAR(r_{ij}) - N_0}{VAR[x_{ij}]}} \quad (2)$$

After a successful decoding, all the symbols of the pass are in hand, ready to update the gain estimator using a better estimator exercising Least Squares, in a way that the subtraction of that pass will be accurate for the successful decoding of the remaining users.

## B. Phase Estimation

Since gain $\alpha_i$ is complex we also need phase estimation for each slot during decoding a specific pass $r_i$. Since we want to use all the data we have, we have added an ancestors tree to the decoder to keep all symbols path up to the current one. The estimator uses the symbols in the current hypothesis during the decoding. Let $j$ be the current symbol in the decoder,

$$\angle \hat{\alpha}_i = \angle \sum_{j=1}^{m}(x_{ij} \cdot r_{ij}^*) \quad (3)$$

The idea is to correlate each candidate with the incoming symbol, and take the phase of the result.

## C. Starting Point Estimation

The Spinal decoder for a certain user requires adding the cost metrics over all passes of a packet. Thus, the timeslot of the first pass of this user need to be identified. We call it the Start timeslot. The Start timeslot is identified simply by a raise in the energy of X dB relative to the thermal noise. The threshold can be chosen so that the false alarm rate is around 0.01 or less. Passing the threshold in an error is not detrimental. having a wrong Start pass will add some noise and may require few additional passes affecting insignificantly the overall rate. The Start timeslot should be determined independently for each user since each user starts at its own time. In the beginning the threshold is passed and the same start timeslot is assigned to all users, but after a user is subtracted, a search for the Start timeslot is carried again and the result is assigned to the currently not decoded users.

## V. NUMERICAL RESULTS

We used a new efficient implementation of the Spinal code

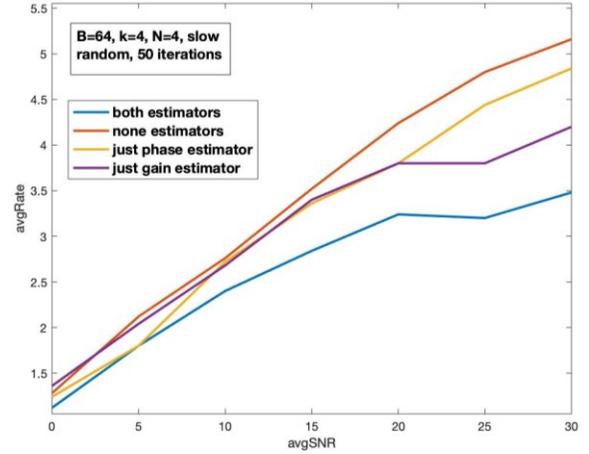

Figure 4: SPIRAP estimators influence comparison

[6] to run the simulation at k=8, B=256 in reasonable run time without puncturing. Puncturing as suggested in [3] is problematic in our setting since we want each user to send the same packets size.

### A. Estimators influence

We evaluate the estimators influence, as described in Figure 4. As can be seen, the phase estimators follows the case of no estimators quite well, while there is more work to be done to improve the case of joint estimators.

### B. User power influence - analysis for two users

The simulation of two users which constantly transmit ($\gamma = 1$). We also compare the performance to ideal TDMA. For a fair comparison we assume that in TDMA each user at its own slot transmit Spinal code, so actually TDMA is equivalent to separated $N$ single users that each have $1/N$ of

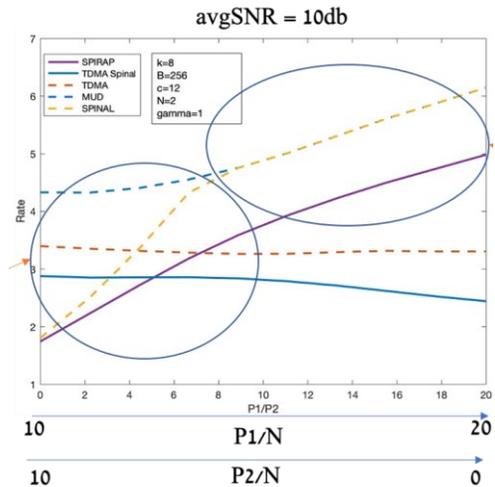

Figure 5: Rate vs P1/P2 SPIRAP bounds

the rate of single user. Since we sum the total average rate, the rate for TDMA equals to the rate of one user transmitting Spinal Code. The TDMA results are also benchmark for

CSMA results, but in reality CSMA performance will be much poorer due to overhead and the use of rateless codes as discussed above.

All curves with dotted lines are theoretical Shannon bounds.

- (Multi User Dectection) MUD theoretical bound: $\log_2(1 + \frac{P1+P2}{N})$
- TDMA bound: $\frac{1}{2}\log_2\left(1 + \frac{P1}{N}\right) + \frac{1}{2}\log_2(1 + \frac{P2}{N})$
- SPIRAP bound
$$\begin{cases} 2\log_2\left(1 + \frac{P1}{P2+N}\right) & ; \log_2\left(1 + \frac{P1}{P2+N}\right) < \log_2\left(1 + \frac{P2}{N}\right) \\ \log_2\left(1 + \frac{P1}{P2+N}\right) + \log_2\left(1 + \frac{P2}{N}\right) & ; else \end{cases}$$

In the working point marked by the left circle drawn on Figure 5, Both users have close power, if one is decoded, the second has high SNR and in high probability will be decoded too. In the working point marked by the right circle, After the strong user is decoded, the weak user is decoded in its own SNR

### C. Random transmissions

Next, we simulate the case that users generate new packets randomly using Poisson distribution parameter $\gamma$. i.e. in each time-slot, each user starts a new packet with probability $\gamma$. We expected to find an optimal $\gamma$ value, since as long $\gamma$ is smaller the average rate is smaller but there are less collisions. The dashed line curves show results for slower fading process than that in the smooth lines cases. We run this simulation for five users, as described next. According to our simulation for 5 users at average SNR of 10dB, best performance are achieved when $\gamma$ value is 0.5 as can be seen from Figure 6. Fast Rayleigh fading channel causes a situation where each user has different gain in each pass, so on average, the users tend to have equal gains. The case of equal gains is the worst for SPIRAP as can be seen in Figure 5. Moreover, higher $k$ means more passes (lower rate) to decode Using $\gamma$, as it increases collision probability.

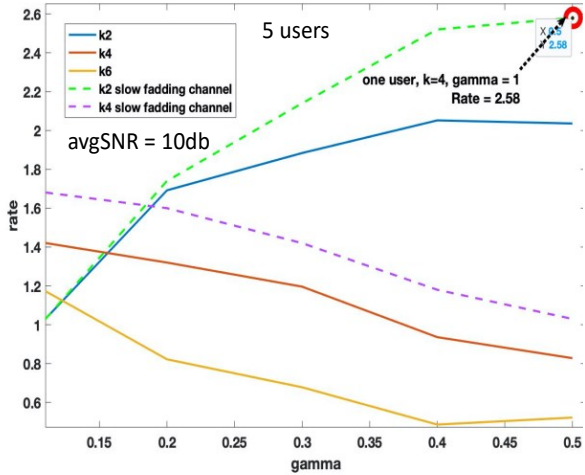

Figure 6: Rate per $\gamma$ per k. 5 users, one user simulate TDMA

The next 3 figures show the influence of the Spinal parameter $k$.

In Figure 7 the evaluation of different $k$'s for 2 users system is demonstrated. In this situation $k$ equals to 8 is optimal.

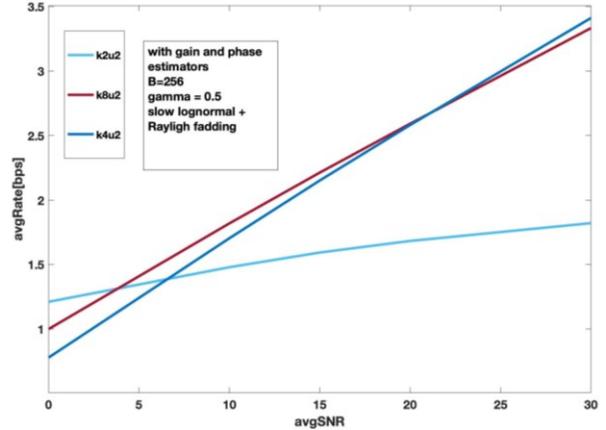

Figure 7: k=2,4,8, number of users = 2

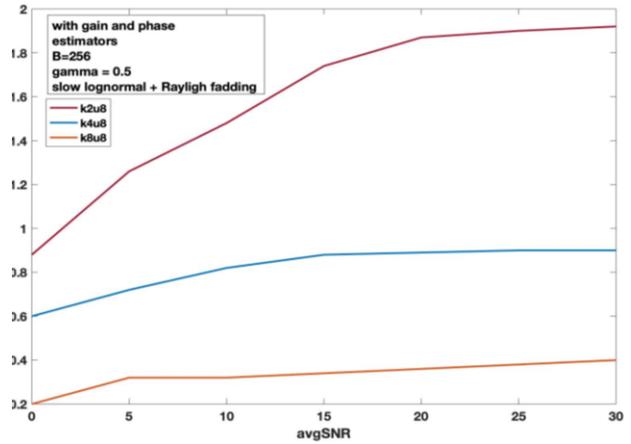

Figure 8: k=2,4,8, number of users = 8

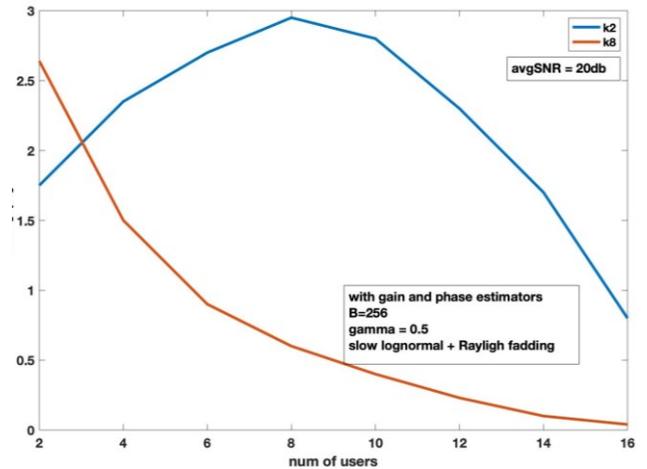

Figure 9: const. k vs. number of users

Next, we simulated the same *k* values again, now for 8 users system as can be seen in figure 8. Thus, in this case *k* =2 is optimal.

The conclusion of this analysis is described in figure 9, where we simulated the rate performance for two different *k* values for incremented number of users. The reason for this result is that for the same SNR when *k* is larger the same average rate is achieved with larger number of passes. When there are more passes the number of collisions increases and as a result, the first (strongest) user has to work in lower SNR.

## VI. SPIRAP VS. CURRENT MULTIPLE ACCESS PROTOCOLS

As described in figure 5, if there is a significant power difference between the users, like the case of near and far users or there is one strong user and one weak, SPIRAP may perform better than TDMA. In real life TDMA has many disadvantages as it can be used only in a tightly controlled system like cellular networks.

As discussed above there are difficulties using rateless codes in the framework of CSMA and ALOHA. In addition to these difficulties, CSMA has some disadvantages compared to SPIRAP as follows:
- Users need to hear each other for synchronization
- Large overhead due to collision avoidance mechanism and packet headers
- Packet headers and control packets need to be transmitted in very low data rate to guarantee the reception in all the range of signal power.

ALOHA has some disadvantages comparing to SPIRAP:
- Low throughput due to collisions and large back-off times
- Poor Fairness, if strong and weak users colide, weak users suffer more.
- Large and unpredictable delay

Moreover, ALOHA type of protocols with spinal code, will have much lower rate than the usual ALOHA due to the retransmissions needed by the Spinal Code.

We cannot compare to current widespread, NOMA (Non-orthogonal Multiple Access) [7], since it is designed for a fully controlled network.

NORA (Non-orthogonal Random Access), which is NOMA based as described in [8], might resemble SPIRAP in the sense that in both protocols the collisions are resolved by successive cancelation. However, there is a profound difference. NORA is based on grants and handshakes, where the base station identifies each user and control what power to transmit in according to the specific channel conditions. NORA requires strict power estimation unlike the rateless code concept.

NORA has NOMA power allocation concept, with centralized management. Users identification is based on a spreading code assuming that users can be resolved by random transmission time. This method cannot work in multipath and requires a large preamble. Some of NORA disadvanatges comparing to SPIRAP are:
- Central management
- The proposed users identification method is problematic and proned to multipath.
- Power allocation based on channel measurements and feedback cannot work in fast mobile channel.
- In a situation of two far users, lowering the power of one of them might cause un-detection.

## VII. CONCLUSION

SPIRAP advantages conclusion:
- SPIRAP achieves good results when using a minimally managed network, no carrier sense, users transmit at will
- No power control needed
- Low latency, minimum handshake
- Good fairness, strong users will be decoded and removed allowing others to be decoded.
- Works in any unpredictable and time varying signal to noise ratio
- Reduces wireless overhead by eliminating the need for headers, using sequential decoding, and pilot using inter-decoder channel estimator.
- Increased throughput thanks to simultaneous decoding of multiple users.
- Surpasses ALOHA and even the ideally managed TDMA when users have different powers.
- All rateless codes are suitable with SPIRAP, Spinal code was chosen in this work as it's the newest one developed with the smallest gap to capacity exsits.